\documentclass[11pt]{article}
\usepackage[T1]{fontenc}
\usepackage[utf8]{inputenc}
\usepackage[english]{babel}
\usepackage{lmodern}
\usepackage{amsmath,amssymb}
\usepackage{dsfont}
\usepackage[a4paper,margin=1in]{geometry}
\usepackage{microtype}
\usepackage{xcolor}
\usepackage[colorlinks=true,allcolors=blue!60!black]{hyperref}
\begin{document}

\title{Fixed Points of the Yang--Mills Gradient Flow\\ in the Vortex Sector}
\author{A. M. Silva\\[4pt] {\small\itshape Instituto de F\'isica, Universidade de S\~ao Paulo (USP),}\\ {\small\itshape S\~ao Paulo 05314-970, Brazil}}
\date{}
\maketitle

\begin{abstract}
We study the Yang--Mills gradient flow in the sector of center vortex configurations in SU(2) gauge theory. For axially symmetric vortices the flow reduces exactly to a linear diffusion equation for the radial profile, which we solve in closed form: a Gaussian profile keeps its shape while the core width grows as $\sqrt{R^2+4t}$, the magnetic flux is exactly conserved, and the thin vortex is the unique bounded fixed point, although not an attractor. Within this Abelian sector the flow has no growing modes. Off-diagonal perturbations are a different story. Their evolution is governed by a pair of radial Schr\"odinger operators with a chromomagnetic potential set by the vortex field, and an explicit variational computation for a Gaussian core, keeping the full gauge barrier, shows that the operator in the dangerous helicity channel has a negative eigenvalue for every winding number $n \geq 1$, with a growth rate approaching the peak chromomagnetic field at large $n$. Because the background spreads while the mode grows, the two effects compete, and we settle the competition exactly: in self-similar variables the unstable modes are known in closed form, with growth exponents $\mu_{n,m}=n-1+m$ in angular sector $m\leq 0$. Multiply wound vortices are therefore unstable at every $n\geq 2$, while the fundamental vortex is exactly marginal and keeps its Abelian form. A scaling argument rules out any static vortex, Abelian or not, in the transverse plane, and the instability acquires a simple geometric meaning: it shortens the path traced in SU(2) by the holonomy of a growing loop, lowering the winding by two at each step. The flow thus drives every vortex toward the Abelian core of winding $n \bmod 2$, with non-Abelian structure appearing as the transient through which multiply wound cores unwind. \par\medskip\noindent{\raggedright\textbf{Keywords:} center vortices, Yang--Mills gradient flow, fixed points, stability analysis, Nielsen--Olesen instability\par}
\end{abstract}

\section{Introduction}
\label{introduction}

The mechanism responsible for confinement in pure Yang--Mills theory remains one of the central open problems in the infrared structure of non-Abelian gauge theories. Among the proposals, the center vortex picture is one of the most compelling candidates for the long-distance dynamics~\cite{tHooft:1977nqb,Cornwall:1979hz}: confinement arises from the percolation of extended topological objects whose linking with Wilson loops generates an area law. Lattice studies provide extensive evidence that center vortices capture key infrared features, including the string tension, topological charge, and aspects of chiral symmetry breaking~\cite{DelDebbio:1996mh,deForcrand:1999ms,Engelhardt:1999fd,Greensite:2003bk}, and a comprehensive account is given in~\cite{Greensite:2016pfc}. The picture belongs to a broader family of scenarios treating topological gauge configurations as the relevant infrared degrees of freedom, alongside Abelian monopoles~\cite{tHooft:1981ht} and instantons~\cite{Belavin:1975fg,Callan:1977gz}, with the vortex--monopole relation studied both on the lattice~\cite{deForcrand:2000pg} and in the continuum~\cite{Reinhardt:2001kf}.

Center vortices are characterized by gauge fields whose holonomy around nontrivial loops takes values in the center of the gauge group. For the case of SU(2), the relevant center is $\mathbb{Z}_2$, and vortex configurations are classified by the nontrivial element of
\begin{equation}
\pi_1(SU(2)/\mathbb{Z}_2) = \mathbb{Z}_2.
\label{eq:pi1_intro}
\end{equation}

Thin vortices can be described as pure gauge configurations everywhere except along a singular core carrying quantized magnetic flux. More realistic physical configurations correspond to thick vortices, where the flux is distributed smoothly over a finite region. Such objects arise naturally in lattice gauge theory after smoothing procedures and play a central role in center-projection approaches to confinement~\cite{Langfeld:1997jx,Faber:2001zs}.

A powerful tool for analyzing the structure of gauge fields is the Yang--Mills gradient flow, which evolves gauge configurations along a fictitious flow time according to the steepest descent of the Yang--Mills action. The flow equation is
\begin{equation}
\partial_t A_\mu = D_\nu F_{\nu\mu}.
\label{eq:YMflow_intro}
\end{equation}
Introduced as a smoothing and renormalization tool for lattice gauge theory~\cite{Narayanan:2006rf,Luscher:2009eq,Luscher:2010iy}, the gradient flow is now standard for defining renormalized observables, setting physical scales, and studying topological properties~\cite{Luscher:2013vga,Borsanyi:2012zs}. Its perturbative properties are also well understood: the flowed fields require no further renormalization, and the small-flow-time expansion is under control~\cite{Luscher:2011bx,Luscher:2013cpa,Suzuki:2013gza}. Because ultraviolet fluctuations are suppressed while long-distance structure persists over a finite range of flow time, infrared objects such as topological charge, instantons and center vortices are expected to survive the smoothing.

Despite this, little is known about how center vortices themselves evolve under the flow. Lattice studies typically smooth generic configurations, by cooling or by the flow itself~\cite{Bonati:2014tqa,Alexandrou:2015yba}, and identify vortices by center projection~\cite{Trewartha:2015ida,Biddle:2019gke}; the intrinsic flow dynamics of vortex configurations, and in particular their stability in the continuum, have not been systematically analyzed.

In this work we investigate the behavior of center vortex configurations under the Yang--Mills gradient flow in SU(2) gauge theory. We focus on axially symmetric vortex configurations whose gauge field is diagonal in color space and depends on a radial profile function. Within this ansatz, the vortex flux lies in the $\sigma_3$ color direction, leading to an effectively Abelian structure for the background configuration.

Our first result is that the Yang--Mills flow equation reduces exactly to a linear diffusion equation for the vortex profile function,
\begin{equation}
\partial_t f = f'' - \frac{1}{r}f' = r\,\partial_r\!\left(\frac{1}{r}\partial_r f\right).
\label{eq:diffusion_intro}
\end{equation}
The relative sign of the first-derivative term is fixed by the one-form character of the ansatz: $f$ is the component of $A_\mu$ in the coordinate basis, not a scalar, so the relevant operator is the vector Laplacian rather than the scalar one. Equivalently, the physical azimuthal component $f/r$ obeys the radial heat equation in the $|m|=1$ sector, and each Cartesian component of $A_\mu$ obeys the ordinary two-dimensional heat equation. This provides a transparent analytical description of how vortex profiles evolve, and for a Gaussian core the flow is solvable in closed form: the profile keeps its shape with a width growing as $\sqrt{R^2+4t}$, the flux is exactly conserved, and both the regularity condition at the origin and the holonomy at infinity are preserved for all flow times. The bounded fixed points are the constant profiles, which reproduce the thin vortex; this configuration is a fixed point but not an attractor, since regular initial data spread diffusively and move away from it.

However, the Abelian ansatz does not capture the full dynamics of the non-Abelian gauge theory. To address this, we analyze perturbations in the off-diagonal color directions. Linearizing the flow equation around the Abelian vortex background leads to an evolution equation for the off-diagonal modes that takes the form of a Schrödinger-type operator with a chromomagnetic potential determined by the vortex magnetic field. In particular, the stability of the Abelian vortex reduces to the spectral properties of the pair of radial operators
\begin{equation}
\mathcal{H}^\pm = -\Delta_{\mathbb{R}^2} + \frac{f(r)^2}{r^2} \pm 2B_3(r),
\label{eq:Hpm_intro}
\end{equation}
written here for the lowest angular sector, where $B_3(r)=f'(r)/r$ is the radial chromomagnetic field of the vortex and the two signs correspond to the two helicities of the off-diagonal gluon.

A key observation is that this stability problem is effectively two-dimensional, a setting in which attractive potentials bind unusually easily: an integrable attractive well in two dimensions always supports a bound state, however weak~\cite{Simon:1976un,ReedSimon1978}. That classic result does not apply verbatim here, because the gauge barrier $f^2/r^2$ is not integrable and carries a factor $n^2$ against the $n$ of the magnetic term, so the competition between the two must be settled quantitatively rather than by appeal to a general theorem. We do so variationally, retaining every term, and find that for the operator
\begin{equation}
\mathcal{H}^- = -\Delta_{\mathbb{R}^2} + \frac{f(r)^2}{r^2} - 2B_3(r)
\label{eq:Hminus_intro}
\end{equation}
the lowest eigenvalue is negative for a Gaussian core of any winding number $n\geq 1$. This implies the existence of tachyonic off-diagonal modes in the vortex background: frozen at any instant, the Abelian core is unstable. Bound states of this kind were already seen numerically, for a range of profiles, in one-loop computations of the vacuum energy of a color-magnetic vortex~\cite{Bordag:2002sa}; what the flow adds is a dynamical meaning for them, and what the Gaussian core adds is a closed-form bound.

The physical interpretation of this instability is closely analogous to the Nielsen--Olesen instability of a uniform chromomagnetic field~\cite{Nielsen:1978rm,Savvidy:1977as}. There, the lowest Landau level of an off-diagonal vector boson in a constant background $B_3$ has a tachyonic mode of mass-squared $-gB_3$, signaling the instability of the Savvidy vacuum and motivating the picture of the true vacuum as a disordered chromomagnetic medium~\cite{Ambjorn:1980ms}. In the present case, however, the instability arises in an inhomogeneous and spatially localized chromomagnetic background generated by the vortex. Rather than a continuous spectrum of Landau levels filling all of space, one obtains a discrete bound state of an attractive radial potential well, localized within the vortex core. The growth of the unstable mode drives the system away from the Abelian ansatz, and the vortex core dynamically develops non-Abelian structure as the flow progresses.

Because the background itself evolves, the instability has to be weighed against the Abelian dilution rather than against a static configuration. The growth rate is set by the peak chromomagnetic field and therefore decays exactly as fast as the core spreads, so the two effects compete on equal terms at all times. We settle the competition exactly. In self-similar variables, in which the spreading Gaussian vortex is stationary, the unstable modes can be written in closed form, and their growth exponents are integers: $\mu_{n,m}=n-1+m$ in angular sector $m\leq 0$, the off-diagonal amplitude growing relative to the background as $(L/R)^{\mu}$ with $L$ the core width. Every multiply wound vortex is therefore unstable, while for the fundamental vortex the frozen-core instability is exactly cancelled by the dilution. Since the center of $\mathrm{SU}(2)$ is $\mathbb{Z}_2$, only the parity of $n$ is topologically protected, and the instability is precisely the channel through which that freedom is used.

This leads to a clean dynamical picture. In the transverse plane the action is scale covariant, and Derrick's argument~\cite{Derrick:1964ww} rules out any static vortex carrying a center charge, Abelian or not; the flow has nowhere to rest, and its late stages are described by self-similar, expanding configurations. Within the symmetric sectors a vortex is a curve in the Lie algebra whose action is a weighted length, and the unstable mode swings that curve about the origin, shortening the path traced in $\mathrm{SU}(2)$ by the holonomy of a growing loop and lowering the winding by two. The flow therefore carries every vortex toward the Abelian core of winding $n\bmod 2$. Non-Abelian structure appears along the way, but as the transient through which multiply wound cores unwind rather than as a new kind of core. This is the planar, non-compact counterpart of the Morse theory of the Yang--Mills functional on compact surfaces~\cite{Atiyah:1982fa,Rade:1992}, with the critical points replaced by expanding solutions. It also sharpens the comparison with the non-Abelian vortices of supersymmetric and Yang--Mills--Higgs theories~\cite{Hanany:2003hp,Auzzi:2003fs,Shifman:2004dr}, whose cores are held together by the scale a Higgs field provides. Pure Yang--Mills theory in the plane has no such scale, but at finite temperature the Polyakov loop supplies one, and that is where a genuinely non-Abelian vortex core should be sought.

Beyond its conceptual interest, this analysis has implications for lattice studies of center vortices. The gradient flow is a standard smoothing tool in lattice calculations~\cite{Bonati:2014tqa,Alexandrou:2015yba}, and any study that smooths configurations this way before identifying vortices inherits the effect described here. Fundamental vortices keep their Abelian form under the flow, while multiply wound cores unwind on a flow-time scale set by their width, so vortex observables extracted after extensive smoothing measure the fundamental population rather than the full structure present in the original configuration.

\section{Center Vortex Configuration in SU(2)}
\label{sec:vortices}

In this section we review the vortex backgrounds that will be evolved under the Yang--Mills gradient flow. We work in Euclidean space and adopt polar coordinates $(r,\phi)$ in the plane transverse to the vortex axis. The configurations considered here are axially symmetric and take values in an Abelian, color-diagonal subgroup of SU(2), with flux oriented along the $\sigma_3$ direction.

\subsection{Thin Vortex}
\label{subsec:thin}

A thin center vortex of winding number $n\in\mathbb{Z}$ is described by the gauge potential
\begin{equation}
A_\mu(x)= n\,\partial_\mu \phi(x)\,\frac{\sigma_3}{2},
\label{eq:thin_vortex_A}
\end{equation}
where $\phi$ is the azimuthal angle and $\sigma_3$ is the third Pauli matrix. Away from the vortex core this field is locally pure gauge. Indeed, for $r\neq 0$ one may write
\begin{equation}
A_\mu = i\, U\,\partial_\mu U^{-1},
\qquad
U(x)=\exp\!\left(i\,n\,\phi(x)\,\frac{\sigma_3}{2}\right),
\label{eq:thin_vortex_pure_gauge}
\end{equation}
so that the field strength vanishes pointwise outside the core,
\begin{equation}
F_{\mu\nu}(x)=0,
\qquad (r\neq 0).
\label{eq:thin_F_zero}
\end{equation}
Nevertheless the configuration is not globally gauge equivalent to the vacuum due to the multivaluedness of $\phi$ at the origin. The associated holonomy around a loop $C$ encircling the origin is
\begin{equation}
W(C)=\mathcal{P}\exp\!\left(i\oint_C A_\mu\,dx^\mu\right)
= \exp\!\left(i\pi n\,\sigma_3\right) = (-1)^n\,\mathds{1},
\label{eq:thin_holonomy}
\end{equation}
which for odd $n$ is the nontrivial center element $-\mathds{1}$ of SU(2). In particular, the vortex is classified by the nontrivial element of the fundamental group~\eqref{eq:pi1_intro} of the adjoint group $SU(2)/\mathbb{Z}_2$. The magnetic flux is therefore concentrated at the core and is distributional in the continuum description. In the thick-vortex construction below, this singularity is regulated by a smooth profile.

\subsection{Thick Vortex}
\label{subsec:thick}

A thick or smeared vortex is obtained by introducing a radial profile $f(r)$ that interpolates between regular behavior at the origin and the correct asymptotic holonomy at large radius:
\begin{equation}
A_\mu(x)= f(r)\,\partial_\mu \phi(x)\,\frac{\sigma_3}{2}.
\label{eq:thick_vortex_A}
\end{equation}
Regularity at $r=0$ requires that $A_\mu$ be nonsingular. Concretely, $f(r)$ must vanish at the origin at least as $\mathcal{O}(r^2)$, so that
\begin{equation}
f(r)\xrightarrow[r\to 0]{} 0.
\label{eq:f_bc_origin}
\end{equation}
The asymptotic value of the profile is fixed by the center holonomy. For a vortex of winding $n$ we impose
\begin{equation}
f(r)\xrightarrow[r\to \infty]{} n.
\label{eq:f_bc_infty}
\end{equation}
With these boundary conditions the configuration carries the same $\mathbb{Z}_2$ center charge as the thin vortex, but with a finite-width core.

Because the background is diagonal in color space, commutator terms vanish and the field strength is purely Abelian. In polar coordinates, the only nonzero component is
\begin{equation}
F_{r\phi}(r)=\partial_r A_\phi
= f'(r)\,\frac{\sigma_3}{2},
\label{eq:Frphi}
\end{equation}
corresponding to a color magnetic field along $\sigma_3$,
\begin{equation}
B_3(r)=\frac{1}{r}\,f'(r).
\label{eq:B3_def}
\end{equation}

The thin vortex is recovered as the limiting case where the profile approaches a step at the origin while maintaining the asymptotic holonomy~\eqref{eq:f_bc_infty}. In particular, the condition~\eqref{eq:f_bc_infty} ensures that at large radius the Wilson loop holonomy matches that of the thin vortex.
\subsection{Example Profiles}
\label{subsec:profiles}

For explicit calculations it is convenient to consider smooth model profiles. A commonly used choice is a Gaussian-type interpolation,
\begin{equation}
f(r)= n\left(1-e^{-r^2/R^2}\right),
\label{eq:gaussian_profile}
\end{equation}
where $R$ sets the core thickness. This yields a localized chromomagnetic field
\begin{equation}
B_3(r)=\frac{1}{r}f'(r)=\frac{2n}{R^2}e^{-r^2/R^2}.
\label{eq:gaussian_B3}
\end{equation}
As an analytically convenient alternative one may also employ sharp profiles, such as step-like interpolations, that reproduce the same holonomy but concentrate the field strength near $r\sim R$. Smooth profiles are physically motivated by lattice center-vortex identification procedures, where vortex cores appear as extended objects once ultraviolet fluctuations are suppressed by smoothing.

In the following sections we will evolve these backgrounds under the Yang--Mills gradient flow. We first treat the color-diagonal sector, where the dynamics reduces to an exact diffusion equation for $f$, and then include off-diagonal perturbations to determine the stability of the Abelian ansatz.

\section{Yang--Mills Flow for the Abelian Sector}
\label{sec:abelian}
We now analyze the Yang--Mills gradient flow for the thick center vortex configuration introduced in Section~\ref{sec:vortices}. The flow equation
\begin{equation}
    \frac{\partial A_\mu}{\partial t} = D^\nu F_{\nu\mu}
    \label{eq:YMflow}
\end{equation}
governs the evolution of the gauge field as a function of the flow time $t \geq 0$. Throughout we use Hermitian generators $T_a=\sigma_a/2$ satisfying $[T_a,T_b]=i\epsilon_{abc}T_c$, write $A_\mu=A^a_\mu T_a$ with real components $A^a_\mu$, and adopt the conventions
\begin{equation}
    F_{\mu\nu} = \partial_\mu A_\nu - \partial_\nu A_\mu - i\,[A_\mu,A_\nu],
    \qquad
    D_\mu X = \partial_\mu X - i\,[A_\mu,X]
    \label{eq:conventions}
\end{equation}
for the field strength and for the covariant derivative acting on an adjoint-valued field $X$, so that in components $(D_\mu X)^c=\partial_\mu X^c+\epsilon_{abc}A^a_\mu X^b$ is manifestly real. The gauge coupling has been absorbed into $A_\mu$. In this section we restrict attention to the color-diagonal sector, in which the gauge field takes the form
\begin{equation}
    A_\mu(x,t) = f(r,t)\,\partial_\mu\phi\;\frac{\sigma_3}{2},
    \label{eq:ansatz}
\end{equation}
where $\phi = \arctan(x^2/x^1)$ is the azimuthal angle in the $(x^1, x^2)$ plane and $r = \sqrt{(x^1)^2+(x^2)^2}$ is the transverse radius. The profile function $f(r,t)$ encodes the core structure of the vortex. The initial profile is required to be regular at the origin, vanishing at least as $\mathcal{O}(r^2)$:
\begin{equation}
    f(r,0) \xrightarrow[r\to 0]{} 0,
    \label{eq:ic_regular}
\end{equation}
while the asymptotic value, which fixes the topological sector, is preserved by the flow,
\begin{equation}
    f(r,t) \;\longrightarrow\; n \quad \text{as } r \to \infty,
    \quad t \geq 0,
    \label{eq:bc}
\end{equation}
for a vortex of winding number $n \in \mathbb{Z}$. Both conditions are preserved by the flow: the exact solution constructed in Section~\ref{sec:heat_kernel} satisfies $f(0,t)=0$ and $f(\infty,t)=n$ for every $t\geq 0$, so a smooth thick vortex remains smooth and stays in its topological sector under the evolution.

\subsection{Flow Equation for the Profile}
\label{sec:profile_flow}

We work in the $(x^1, x^2)$ plane and use polar coordinates $(r, \phi)$. The only non-vanishing component of the gauge field in these coordinates is
\begin{equation}
    A_\phi = f(r,t)\,\frac{\sigma_3}{2}, \qquad A_r = 0.
    \label{eq:polar_components}
\end{equation}
The associated field strength tensor has a single independent component,
\begin{equation}
    F_{r\phi} = \partial_r A_\phi - \partial_\phi A_r - i\,[A_r, A_\phi]
              = f'(r,t)\,\frac{\sigma_3}{2},
    \label{eq:field_strength}
\end{equation}
where primes denote $\partial_r$, and the commutator vanishes identically since both $A_r$ and $A_\phi$ are proportional to $\sigma_3$. The corresponding magnetic field in the transverse plane is $F_{12}=F_{r\phi}/r = B_3\,\sigma_3/2$, with
\begin{equation}
    B_3(r,t) \;=\; \frac{f'(r,t)}{r},
    \label{eq:magnetic}
\end{equation}
in agreement with~\eqref{eq:B3_def}.

We now reduce the flow equation~\eqref{eq:YMflow} to an equation for $f$. Since the configuration is color-diagonal, the covariant derivative reduces to the ordinary derivative,
\begin{equation}
    D^\nu F_{\nu\mu} = \partial^\nu F_{\nu\mu} - i\,[A^\nu, F_{\nu\mu}]
    = \partial^\nu F_{\nu\mu},
    \label{eq:covderiv_abelian}
\end{equation}
because every commutator is of the form $[\sigma_3, \sigma_3] = 0$. The problem is therefore Abelian in color space, and the flow equation reduces to a single scalar equation for the profile function.

Evaluating $\partial^\nu F_{\nu\phi}$ requires some care, because $f$ is the component of a one-form in the coordinate basis rather than a scalar. With the transverse metric $\mathrm{d}s^2=\mathrm{d}r^2+r^2\mathrm{d}\phi^2$ and $\sqrt{g}=r$, the divergence of an antisymmetric tensor is $\nabla_\nu F^{\nu\mu}=g^{-1/2}\partial_\nu(g^{1/2}F^{\nu\mu})$. Raising both indices gives $F^{r\phi}=g^{rr}g^{\phi\phi}F_{r\phi}=f'/r^2$, so
\begin{equation}
    \nabla_\nu F^{\nu\phi}
    = \frac{1}{r}\,\partial_r\!\left(r\,\frac{f'}{r^2}\right)
    = \frac{f''}{r^2} - \frac{f'}{r^3},
    \label{eq:divup}
\end{equation}
and lowering the free index with $g_{\phi\phi}=r^2$,
\begin{equation}
    \partial^\nu F_{\nu\phi} = r^2\,\nabla_\nu F^{\nu\phi}
    \;=\; f''(r,t) - \frac{f'(r,t)}{r}.
    \label{eq:laplacian_f}
\end{equation}
Substituting into~\eqref{eq:YMflow} and projecting onto the $\sigma_3$ direction, we obtain the evolution equation for the profile:
\begin{equation}
    \frac{\partial f}{\partial t}
    \;=\; f'' - \frac{f'}{r}
    \;=\; r\,\partial_r\!\left(\frac{1}{r}\,\partial_r f\right).
    \label{eq:profile_flow}
\end{equation}

Equation~\eqref{eq:profile_flow} is the central result of this section. It is a diffusion equation, but not the two-dimensional radial heat equation for $f$: the relative sign of the first-derivative term reflects the one-form character of the ansatz. Its structure is clearest in Cartesian components, where $A_x=-f\,y/r^2$ and $A_y=f\,x/r^2$ each obey the ordinary two-dimensional heat equation $\partial_t A_i=\Delta A_i$, the Coulomb-gauge condition $\partial_iA_i=0$ being satisfied identically by the axially symmetric azimuthal ansatz. Equivalently, the physical azimuthal component $a=A_{\hat\phi}=f/r$ obeys $\partial_t a = a''+a'/r-a/r^2$, the radial heat equation in the $|m|=1$ sector, the last term arising from the rotation of the azimuthal basis vector. The Yang--Mills flow therefore acts on the vortex as ordinary diffusion of the gauge potential in the transverse plane.

The fact that the non-Abelian commutator $[A^\nu, F_{\nu\mu}]$ vanishes for this ansatz is not accidental. It is a direct consequence of the color-diagonal structure of~\eqref{eq:ansatz}: all gauge field components are proportional to the same generator $\sigma_3/2$, so they commute among themselves. As we will see in Section~\ref{sec:offdiag}, this simplification breaks down as soon as off-diagonal $\sigma_1,\sigma_2$ perturbations are introduced, and the full non-Abelian structure of the flow is activated. The analysis of the present section therefore applies strictly within the Abelian sector, and its results should be interpreted as governing the longitudinal color-diagonal dynamics of the vortex.

The thin vortex corresponds to the profile
\begin{equation}
    f_{\rm thin}(r) = n, \qquad r > 0,
    \label{eq:thin_profile}
\end{equation}
which is a constant in $r$ away from the singularity at the origin. Substituting into~\eqref{eq:profile_flow} gives $\partial_t f_{\rm thin} = 0$ identically: the thin vortex is a fixed point of the flow. This reproduces from the profile equation the result obtained earlier by direct inspection of the field strength, namely that $F_{\mu\nu} = 0$ for $r>0$ implies $D^\nu F_{\nu\mu} = 0$. The fixed-point structure in the Abelian sector is analyzed in Section~\ref{sec:fixedpoints}. We emphasize already here that $f_{\rm thin}$ is a fixed point but not an attractor: as the exact solution of Section~\ref{sec:heat_kernel} shows, regular initial data flow away from it.

\subsection{Fixed Points}
\label{sec:fixedpoints}
A fixed point of the flow~\eqref{eq:profile_flow} is a time-independent profile $f_*(r)$ satisfying
\begin{equation}
    r\,\partial_r\!\left(\frac{1}{r}\,\partial_r f_*\right) = 0,
    \qquad r > 0.
    \label{eq:fixed_point_eq}
\end{equation}
Integrating twice gives the general solution
\begin{equation}
    f_*(r) = \beta + \alpha\, r^2,
    \label{eq:general_fixed_point}
\end{equation}
with $\alpha,\beta \in \mathbb{R}$. The two constants have a transparent meaning: $\beta$ fixes the holonomy, while $\alpha$ corresponds to a uniform chromomagnetic field $B_3 = f_*'/r = 2\alpha$, which is of course a static solution of the Abelian field equations. Boundedness of the holonomy as $r \to \infty$ requires $\alpha = 0$, and the boundary conditions~\eqref{eq:bc} then give $\beta = n$. Hence
\begin{equation}
    f_*(r) = n,
    \label{eq:unique_fixed_point}
\end{equation}
which is precisely the thin vortex profile~\eqref{eq:thin_profile}. Within the class of profiles satisfying the boundary conditions~\eqref{eq:bc}, the thin vortex $f_* = n$ is therefore the unique fixed point of the Yang--Mills flow~\eqref{eq:profile_flow} in the Abelian sector.

It is essential not to confuse this uniqueness statement with a statement about attraction. The thin vortex carries a distributional magnetic flux concentrated at the origin and is singular there, whereas the flow acting on regular initial data produces a configuration that remains regular for all $t$. As the exact solution of the next subsection shows, a smooth thick vortex does not approach $f_*$; its core spreads diffusively and the profile tends pointwise to zero, with the flux carried out to ever larger radius. The thin vortex is a fixed point from which the flow moves away, not one to which it converges.

\subsection{Exact Solution via the Heat Kernel}
\label{sec:heat_kernel}

Because each Cartesian component of $A_\mu$ obeys the ordinary two-dimensional heat equation, the flow can be solved in closed form by convolution with the heat kernel on $\mathbb{R}^2$~\cite{Grigoryan:2009,Berline:2003}; equivalently, writing $A_y = x\,h(r)$ with $h=f/r^2$ shows that $h$ obeys the radial heat equation in four dimensions, $\partial_t h = h''+3h'/r$, whose static solution $h=n/r^2$ reproduces the fixed point $f=n$ found above. For the physically motivated initial profile
\begin{equation}
    f_0(r) = n\!\left(1 - e^{-r^2/R^2}\right),
    \label{eq:gaussian_profile_flow}
\end{equation}
which vanishes quadratically at the origin, approaches $n$ at large $r$, and has a smooth core of characteristic width $R$, no kernel is needed at all. Inserting the self-similar ansatz $f = n\bigl(1-e^{-r^2/S(t)}\bigr)$ into~\eqref{eq:profile_flow} gives $\dot S = 4$, so that $S = R^2+4t$ and
\begin{equation}
    f(r,t) = n\left(1
    - \exp\!\left(-\frac{r^2}{R^2+4t}\right)\right).
    \label{eq:gaussian_solution}
\end{equation}
The Gaussian profile is thus form-invariant under the flow, with an effective width
\begin{equation}
    R_{\rm eff}(t) = \sqrt{R^2 + 4t},
    \label{eq:effective_width}
\end{equation}
that grows diffusively as $R_{\rm eff} \sim 2\sqrt{t}$ for $t \gg R^2$. Both boundary conditions are preserved exactly: $f(0,t)=0$ for all $t$, so the configuration remains regular at the origin, and $f(\infty,t)=n$, so the topological sector is unchanged. In particular the magnetic field
\begin{equation}
    B_3(r,t) = \frac{f'(r,t)}{r}
    = \frac{2n}{R^2+4t}\,\exp\!\left(-\frac{r^2}{R^2+4t}\right)
    \label{eq:B3_flow}
\end{equation}
broadens while its peak value decreases as $B_3(0,t)=2n/(R^2+4t)$, and the total magnetic flux
\begin{equation}
    \Phi(t) = 2\pi\int_0^\infty B_3(r,t)\,r\,\mathrm{d}r
    = 2\pi\bigl[f(\infty,t)-f(0,t)\bigr] = 2\pi n
    \label{eq:total_flux}
\end{equation}
is conserved for all $t \geq 0$, as required for a topological quantity. There is no tension between flux conservation and the spreading of the core: the flux is fixed by the boundary values alone, which the flow leaves invariant, while the field strength redistributes over an ever larger area.

At fixed $r$ and large $t$, equation~\eqref{eq:gaussian_solution} gives
\begin{equation}
    f(r,t) \;\approx\; \frac{n\,r^2}{4t},
    \label{eq:late_time}
\end{equation}
so the profile decays as $1/t$ towards zero on any compact region while the radius at which it reaches its asymptotic value recedes as $\sqrt{t}$. Within the Abelian sector the flow therefore dilutes the vortex indefinitely rather than sharpening it; the only scale in the problem is the initial core size, and the flow simply erodes it. The evolution is algebraic in flow time rather than exponential, which is characteristic of the continuous spectrum of the linearized operator discussed in Section~\ref{sec:perturbative_stability}.

\subsection{Perturbative Stability of the Fixed Point}
\label{sec:perturbative_stability}

To analyze the stability of the thin vortex as a fixed point of the flow~\eqref{eq:profile_flow}, we write
\begin{equation}
    f(r,t) = n + \varepsilon\, g(r,t)
    \label{eq:perturbation}
\end{equation}
and expand to first order in $\varepsilon$. Since~\eqref{eq:profile_flow} is linear, $g$ satisfies exactly the same equation,
\begin{equation}
    \frac{\partial g}{\partial t}
    = r\,\partial_r\!\left(\frac{1}{r}\,\partial_r g\right).
    \label{eq:perturbation_eq}
\end{equation}
We seek normal-mode solutions of the form $g(r,t) = e^{-\lambda t}\,g_\lambda(r)$. Substituting gives the eigenvalue problem
\begin{equation}
    -\,r\,\partial_r\!\left(\frac{1}{r}\,g_\lambda'\right) = \lambda\, g_\lambda .
    \label{eq:eigenvalue_problem}
\end{equation}
The operator on the left is symmetric with respect to the measure $\mathrm{d}r/r$ rather than $r\,\mathrm{d}r$, and its quadratic form is
\begin{equation}
    \bigl\langle g, -r\partial_r(r^{-1}\partial_r) g\bigr\rangle_{\mathrm{d}r/r}
    = \int_0^\infty \frac{\bigl(g'(r)\bigr)^2}{r}\,\mathrm{d}r \;\geq\; 0 ,
    \label{eq:quadform}
\end{equation}
so the spectrum is non-negative. Equivalently, writing $g = r\,\gamma$ reduces~\eqref{eq:eigenvalue_problem} to the radial Helmholtz equation in the $|m|=1$ sector, whose regular solution at the origin is
\begin{equation}
    g_\lambda(r) = r\,J_1\!\left(\sqrt{\lambda}\,r\right),
    \label{eq:eigenmode}
\end{equation}
with $J_1$ the Bessel function of the first kind of order one. There is no discrete spectrum: the spectrum is purely continuous, $\lambda \in [0,\infty)$, and a general perturbation decomposes as a Hankel transform of order one over these modes.

Since all eigenvalues satisfy $\lambda \geq 0$, every perturbation mode either decays for $\lambda > 0$ or remains constant for $\lambda = 0$ in flow time. The $\lambda = 0$ zero mode, $g_0 = \text{const}$, corresponds to a uniform shift of the profile value and is associated with the freedom to adjust the vortex winding number; it is removed by fixing the boundary condition $f \to n$. The thin vortex fixed point $f_* = n$ is therefore marginally stable under color-diagonal perturbations of the profile: all normal modes decay or are neutral, and no tachyonic, exponentially growing modes exist in the Abelian sector.

The absence of negative eigenvalues is consistent with the diffusive nature of the flow: equation~\eqref{eq:profile_flow} is a well-posed parabolic PDE, and the maximum principle forbids the growth of perturbations with the prescribed boundary conditions. The evolution is algebraic rather than exponential, reflecting the marginal $\lambda = 0$ mode and the fact that the operator has no spectral gap.

It is worth being precise about what this establishes. There is no tachyonic mode in the color-diagonal sector: a thick vortex placed in the Abelian ansatz~\eqref{eq:ansatz} develops no exponentially growing $\sigma_3$ perturbation. But neither does it settle down. The exact solution~\eqref{eq:gaussian_solution} shows that it spreads diffusively, the core width growing without bound and the field diluting, while flux and holonomy stay fixed. The Abelian sector contains no regular stationary configuration at finite core size: the action of the Gaussian family, per unit area of the vortex world-sheet and in units of $1/g^2$, is
\begin{equation}
    S_{\rm YM} = \frac{1}{2}\int \mathrm{d}^2x\,B_3^2 = \frac{\pi n^2}{R_{\rm eff}^2},
    \label{eq:action_scaling}
\end{equation}
which decreases monotonically as the core spreads and diverges in the thin limit. The thin vortex is thus the maximum of the action along this family, not its minimum, and a steepest-descent flow necessarily moves away from it. The Abelian evolution is a scale-dilution runaway driven by the absence of any scale in classical Yang--Mills theory, and the marginal stability established above is stability against tachyonic growth, not against this dilution.

This analysis is, moreover, blind to the off-diagonal $\sigma_1,\sigma_2$ directions in color space. As we demonstrate in Section~\ref{sec:offdiag}, the vortex background $B_3(r,t)$ acts as a chromomagnetic potential for off-diagonal gluon fluctuations, and the situation there is qualitatively different: the relevant operator acquires an attractive potential well, and whether it supports bound states, that is, tachyonic modes, becomes a genuine question. The absence of Abelian growing modes is therefore a necessary but not sufficient condition for stability of the center vortex within the full SU(2) theory.

\section{Off-Diagonal Perturbations and Non-Abelian Stability}
\label{sec:offdiag}

The stability analysis of Section~\ref{sec:perturbative_stability} established that the thin vortex fixed point is marginally stable under color-diagonal perturbations. That analysis was blind to the full color structure of $\mathrm{SU}(2)$: all perturbations were constrained to the $\sigma_3$ direction and non-Abelian commutator terms vanished identically. We now allow fluctuations in all three color directions and show that the off-diagonal sector exhibits a qualitatively different and physically richer behavior. The principal result of this section is that the Abelian thick vortex is unstable under off-diagonal perturbations in the Yang--Mills gradient flow. We establish this by explicit variational computation for the Gaussian core~\eqref{eq:gaussian_profile}, obtaining a closed-form bound valid for every winding number $n \geq 1$, and we then show which features of the argument carry over to other profiles.

We decompose the gauge field as
\begin{equation}
    A_\mu = \bar{A}_\mu + a_\mu, \qquad
    \bar{A}_\mu = f(r)\,\partial_\mu\phi\;T_3, \qquad
    a_\mu = a_\mu^a T_a,
    \label{eq:split}
\end{equation}
with $a_\mu^a\in\mathbb{R}$ and generators as fixed in~\eqref{eq:conventions}. The diagonal component $a^3$ was treated in Section~\ref{sec:abelian}. Here we analyze $a^{1,2}$.

\subsection{The Linearized Flow Operator}
\label{sec:linearized}

Writing $F_{\nu\mu}=\bar{F}_{\nu\mu}+f_{\nu\mu}+\mathcal{O}(a^2)$ with $f_{\nu\mu}=\bar{D}_\nu a_\mu-\bar{D}_\mu a_\nu$, and expanding $\partial_t A_\mu = D^\nu F_{\nu\mu}$ to first order in $a_\mu^{1,2}$ using the conventions~\eqref{eq:conventions}, the perturbation satisfies
\begin{equation}
    \frac{\partial a_\mu}{\partial t}
    = \bar{D}^\nu f_{\nu\mu} - i\,[a^\nu,\,\bar{F}_{\nu\mu}],
    \label{eq:lin_flow}
\end{equation}
where $\bar{D}_\mu a_\nu = \partial_\mu a_\nu - i[\bar{A}_\mu,a_\nu]$. The first term is the covariant kinetic operator; the second is the Pauli coupling, the interaction of off-diagonal gluon spin with the background chromomagnetic field. It is this term that will drive the instability.

Since $\bar{A}_\mu=\bar{A}^3_\mu T_3$, the covariant derivative acts on the off-diagonal components as $(\bar{D}_\mu a_\nu)^c=\partial_\mu a^c_\nu+\epsilon_{3bc}\bar{A}^3_\mu a^b_\nu$. With $\epsilon_{321}=-1$ and $\epsilon_{312}=+1$ this gives
\begin{align}
    (\bar{D}_\mu a_\nu)^1 &= \partial_\mu a^1_\nu
      - \bar{A}^3_\mu\,a^2_\nu, \label{eq:D1}\\
    (\bar{D}_\mu a_\nu)^2 &= \partial_\mu a^2_\nu
      + \bar{A}^3_\mu\,a^1_\nu, \label{eq:D2}
\end{align}
which are real, as they must be. This is the structure of a $\mathrm{U}(1)$ covariant derivative acting on a charged doublet: the background $\bar{A}^3_\mu$ plays the role of an Abelian gauge field. Defining
\begin{equation}
    a^\pm_\mu = a^1_\mu \pm i\,a^2_\mu,
    \label{eq:apm}
\end{equation}
equations~\eqref{eq:D1}--\eqref{eq:D2} combine to give
\begin{equation}
    (\bar{D}_\mu a_\nu)^\pm
    = \bigl(\partial_\mu \pm i\bar{A}^3_\mu\bigr)a^\pm_\nu
    \equiv \mathcal{D}^\pm_\mu\,a^\pm_\nu,
    \label{eq:Dpm}
\end{equation}
so $a^\pm_\mu$ transform as charge-$\pm 1$ matter fields under the background $\mathrm{U}(1)$. Since $a^{1,2}_\mu$ are real we have $a^-_\mu=(a^+_\mu)^*$, and it is enough to analyze the charge-$+$ sector.

For the Pauli term, $-i[X,\bar{F}_{\nu\mu}]$ with $\bar{F}_{\nu\mu}=\bar{F}^3_{\nu\mu}T_3$ has components $\epsilon_{a3c}\,a^{\nu a}\bar{F}^3_{\nu\mu}$, so that
\begin{equation}
    \bigl(-i[a^\nu,\bar{F}_{\nu\mu}]\bigr)^1
    = a^{2}_\nu\bar{F}^{3\nu}{}_\mu,\qquad
    \bigl(-i[a^\nu,\bar{F}_{\nu\mu}]\bigr)^2
    = -\,a^{1}_\nu\bar{F}^{3\nu}{}_\mu,
    \label{eq:pauli_comp}
\end{equation}
again real. Forming the combination $(\cdot)^1\pm i(\cdot)^2$ and using $(\mp i)(\pm i)=+1$,
\begin{equation}
    \bigl(-i[a^\nu,\bar{F}_{\nu\mu}]\bigr)^\pm
    = \mp\,i\,\bar{F}^{3\nu}{}_\mu\,a^\pm_\nu.
    \label{eq:pauli}
\end{equation}

We impose background Lorenz gauge $\mathcal{D}^{\mu,\pm}a^\pm_\mu=0$, which removes the pure-gauge longitudinal degree of freedom. The covariant kinetic term is expanded as $\bar{D}^\nu f_{\nu\mu}\big|^\pm=(\mathcal{D}^\nu\mathcal{D}_\nu)^\pm a^\pm_\mu -(\mathcal{D}^\nu\mathcal{D}_\mu)^\pm a^\pm_\nu$. Reordering the derivatives in the second term with the curvature commutator $[\mathcal{D}^\pm_\mu,\mathcal{D}^\pm_\nu]=\pm i\bar{F}^3_{\mu\nu}$ gives $\mathcal{D}_\mu(\mathcal{D}\!\cdot\!a)^\pm \pm i\bar{F}^{3\nu}{}_\mu a^\pm_\nu$, whose first piece vanishes in the Lorenz gauge. The reordering therefore contributes $\mp i\bar{F}^{3\nu}{}_\mu a^\pm_\nu$, identical to the explicit Pauli term~\eqref{eq:pauli}, and the two add to give the factor of $2$:
\begin{equation}
    \frac{\partial a^\pm_\mu}{\partial t}
    = (\mathcal{D}^\nu\mathcal{D}_\nu)^\pm a^\pm_\mu
        \mp 2i\,\bar{F}^{3\nu}{}_\mu\,a^\pm_\nu
    \equiv -\Delta^\pm_{\mathrm{YM}}\,a^\pm_\mu.
    \label{eq:YM_op}
\end{equation}
Strictly speaking, the flow does not preserve the Lorenz condition. The standard remedy~\cite{Luscher:2010iy} is to add the gauge-restoring term $\bar{D}_\mu\bar{D}^\nu a_\nu$ to the right-hand side, which changes the flow only by a flow-time-dependent gauge transformation, leaves gauge-invariant quantities untouched, and turns~\eqref{eq:YM_op} into an exact statement with no constraint on $a_\mu$. Nothing below depends on this choice. Without the extra term, the quadratic form of the linearized flow operator is $\langle a,\Delta_{\mathrm{YM}}a\rangle-\|\bar{D}\cdot a\|^2$, never larger than that of $\Delta_{\mathrm{YM}}$, so any field with $\langle a,\Delta_{\mathrm{YM}}a\rangle<0$ is a growing direction of the unmodified flow as well.

We now specialize to the two-dimensional transverse $(x^1,x^2)$ plane with polar coordinates $(r,\phi)$. The only non-vanishing background field strength is $\bar{F}^3_{12}=B_3(r)=f'(r)/r$, so that $\bar{F}^{3j}{}_i = -B_3(r)\,\epsilon_{ij}$ and the magnetic coupling in~\eqref{eq:YM_op} becomes $\pm 2i B_3\,\epsilon_{ij}a^\pm_j$. Within the charge-$+$ sector the two spatial components combine into helicity eigenstates,
\begin{equation}
    \Psi = a^+_1 + i\,a^+_2, \qquad \Xi = a^+_1 - i\,a^+_2,
    \label{eq:helicity}
\end{equation}
on which $\epsilon_{ij}a^+_j$ acts diagonally. Explicitly, the coupling has components $2iB_3\,a^+_2$ for $i=1$ and $-2iB_3\,a^+_1$ for $i=2$, so
\begin{align}
    \bigl[2iB_3\epsilon_{ij}a^+_j\bigr]_{i=1}
      + i\bigl[2iB_3\epsilon_{ij}a^+_j\bigr]_{i=2}
    &= 2iB_3 a^+_2 + 2B_3 a^+_1 = +2B_3\,\Psi,
    \label{eq:mag_Psi}\\
    \bigl[2iB_3\epsilon_{ij}a^+_j\bigr]_{i=1}
      - i\bigl[2iB_3\epsilon_{ij}a^+_j\bigr]_{i=2}
    &= 2iB_3 a^+_2 - 2B_3 a^+_1 = -2B_3\,\Xi.
    \label{eq:mag_Xi}
\end{align}
The imaginary unit in~\eqref{eq:YM_op} is cancelled by the antisymmetry of $\epsilon_{ij}$, so the chromomagnetic coupling is a real potential of opposite sign for the two helicities, as it must be for a self-adjoint problem. The transverse flow equations are therefore
\begin{equation}
    \frac{\partial\Psi}{\partial t}
    = \bigl[(\mathcal{D}^j\mathcal{D}_j)^+ + 2B_3(r)\bigr]\Psi,
    \qquad
    \frac{\partial\Xi}{\partial t}
    = \bigl[(\mathcal{D}^j\mathcal{D}_j)^+ - 2B_3(r)\bigr]\Xi.
    \label{eq:Psi_eq}
\end{equation}

Setting $\Psi,\Xi \propto e^{im\phi}e^{-\lambda t}\psi(r)$ with $m\in\mathbb{Z}$ and expanding the covariant Laplacian,
\begin{equation}
    (\mathcal{D}^j\mathcal{D}_j)^+e^{im\phi}\psi
    = e^{im\phi}\!\left[\partial_r^2 + \frac{1}{r}\partial_r
      - \frac{(m+f(r))^2}{r^2}\right]\psi,
    \label{eq:covlap}
\end{equation}
the two equations in~\eqref{eq:Psi_eq} become a pair of decoupled radial eigenvalue problems,
\begin{equation}
    \mathcal{H}^\pm_m\,\psi = \lambda\,\psi, \qquad
    \mathcal{H}^\pm_m
    = -\partial_r^2 - \frac{1}{r}\partial_r
    + \frac{(m+f(r))^2}{r^2} \pm 2B_3(r),
    \label{eq:Hpm}
\end{equation}
with $\mathcal{H}^+_m$ governing $\Xi$ and $\mathcal{H}^-_m$ governing $\Psi$. Both are real and self-adjoint on $L^2(\mathbb{R}^+,r\,\mathrm{d}r)$, and no diagonalization of a complex potential is required: the helicity decomposition~\eqref{eq:helicity} has already done the work. The perturbation is unstable if and only if one of $\mathcal{H}^\pm_m$ has a negative eigenvalue for some $m$.

The operator $\mathcal{H}^+_m$ carries the repulsive term $+2B_3(r)>0$ on top of a non-negative barrier and is manifestly positive semi-definite: there is no instability in the $\Xi$ sector. The operator $\mathcal{H}^-_m$ contains the attractive chromomagnetic term $-2B_3(r)\leq 0$, competing with the centrifugal-plus-gauge barrier $(m+f)^2/r^2\geq 0$. The stability of the Abelian vortex reduces to asking whether this competition produces a bound state.

\subsection{Existence of an Unstable Mode}
\label{sec:stability}

We now establish that $\mathcal{H}^-_0$, the $m=0$ sector of the dangerous operator, has a negative eigenvalue for the Gaussian core with any winding $n\geq 1$. The argument is a variational one and proceeds by exhibiting an explicit trial function whose Rayleigh quotient is strictly negative.

For $m=0$, the operator~\eqref{eq:Hpm} reduces to
\begin{equation}
    \mathcal{H}^-_0
    = -\partial_r^2 - \frac{1}{r}\partial_r
    + \frac{f(r)^2}{r^2} - 2B_3(r),
    \label{eq:H0}
\end{equation}
acting on $L^2(\mathbb{R}^+,r\,\mathrm{d}r)$. This is the two-dimensional Schr\"odinger operator~\cite{ReedSimon1978}
\begin{equation}
    H = -\Delta_{\mathbb{R}^2} + W(r), \qquad
    W(r) = \frac{f(r)^2}{r^2} - 2B_3(r),
    \label{eq:H_2d}
\end{equation}
restricted to radially symmetric functions in $L^2(\mathbb{R}^2)$. By the variational principle, $\inf\mathrm{spec}(H)\leq\langle\psi|H|\psi\rangle/\langle\psi|\psi\rangle$ for any nonzero $\psi\in\mathcal{D}(H)$. We exhibit a trial function for which the right-hand side is strictly negative.

Choose
\begin{equation}
    \psi_\sigma(r) = e^{-r^2/(2\sigma^2)}, \qquad \sigma>0.
    \label{eq:trial}
\end{equation}
On $L^2(\mathbb{R}^2)$ with the radial measure, the normalization and kinetic term are:
\begin{align}
    \|\psi_\sigma\|^2
    &= 2\pi\!\int_0^\infty e^{-r^2/\sigma^2}\,r\,\mathrm{d}r
     = \pi\sigma^2,
    \label{eq:norm}\\
    \|\nabla\psi_\sigma\|^2
    &= 2\pi\!\int_0^\infty \frac{r^2}{\sigma^4}\,
       e^{-r^2/\sigma^2}\,r\,\mathrm{d}r
     = \frac{2\pi}{\sigma^4}\cdot\frac{\sigma^4}{2}
     = \pi.
    \label{eq:kinetic}
\end{align}
Here the kinetic integral uses $\int_0^\infty r^3 e^{-r^2/\sigma^2}\mathrm{d}r =\sigma^4/2$.

Both remaining terms are exactly computable for the Gaussian profile $f(r)=n(1-e^{-r^2/R^2})$, which gives $B_3(r)=(2n/R^2)e^{-r^2/R^2}$. It is convenient to introduce the dimensionless width
\begin{equation}
    s = \frac{\sigma^2}{R^2}.
    \label{eq:sdef}
\end{equation}
For the gauge barrier, substituting $u=r^2$ and expanding $(1-e^{-u/R^2})^2$ reduces the integral to a sum of Frullani integrals $\int_0^\infty u^{-1}(e^{-pu}-e^{-qu})\,\mathrm{d}u=\ln(q/p)$, giving the closed form
\begin{equation}
    V_{\rm gauge}
    = 2\pi\!\int_0^\infty \frac{f(r)^2}{r^2}\,
      e^{-r^2/\sigma^2}\,r\,\mathrm{d}r
    = \pi n^2\,\ln\!\frac{(1+s)^2}{1+2s}\;\geq\;0 .
    \label{eq:Vgauge}
\end{equation}
For the magnetic term the Gaussian integral $\int_0^\infty e^{-\alpha r^2}r\,\mathrm{d}r=1/(2\alpha)$ with $\alpha=(\sigma^2+R^2)/(\sigma^2R^2)$ gives
\begin{equation}
    V_{\rm mag}
    = 2\pi\!\int_0^\infty 2B_3(r)\,e^{-r^2/\sigma^2}\,r\,\mathrm{d}r
    = \frac{8\pi n}{R^2}\!\int_0^\infty
      e^{-r^2(R^{-2}+\sigma^{-2})}\,r\,\mathrm{d}r
    = \frac{4\pi n\,s}{1+s}.
    \label{eq:Vmag}
\end{equation}

Retaining $V_{\rm gauge}$ is essential: discarding it would lower the numerator and hence give a lower bound on $\langle H\rangle_\sigma$, from which nothing follows about $\inf\mathrm{spec}(H)$, and in any case $V_{\rm gauge}$ grows as $n^2$ against the $n$ of $V_{\rm mag}$, so the barrier is not a small correction at large winding. Using~\eqref{eq:norm}, \eqref{eq:kinetic}, \eqref{eq:Vgauge} and~\eqref{eq:Vmag}, the Rayleigh quotient follows exactly:
\begin{equation}
    \langle H\rangle_\sigma
    \equiv \frac{\|\nabla\psi_\sigma\|^2 + V_{\rm gauge} - V_{\rm mag}}
                {\|\psi_\sigma\|^2}
    = \frac{1}{R^2}\,G_n(s), \qquad
    G_n(s) = \frac{1}{s}\left[1 + n^2\ln\frac{(1+s)^2}{1+2s}
             - \frac{4ns}{1+s}\right].
    \label{eq:rayleigh_exact}
\end{equation}

To establish instability it suffices to exhibit one width at which $G_n(s)<0$. The choice
\begin{equation}
    s = \frac{1}{n}
    \label{eq:schoice}
\end{equation}
corresponds to $\sigma = R/\sqrt{n}$, the scale on which the chromomagnetic field is concentrated, and is close to optimal: expanding $G_n$ for small $s$ gives $1/s + (n^2+4n)s - 4n$, minimized at $s=(n^2+4n)^{-1/2}$, which approaches $1/n$ at large $n$. Substituting~\eqref{eq:schoice} into~\eqref{eq:rayleigh_exact} and using $(1+1/n)^2/(1+2/n)=(n+1)^2/[n(n+2)]$ gives $G_n(1/n)=n\,\Lambda(n)$ with
\begin{equation}
    \Lambda(n) = 1 + n^2\ln\!\frac{(n+1)^2}{n(n+2)} - \frac{4n}{n+1}.
    \label{eq:Lambda}
\end{equation}
The logarithm is bounded using $\frac{(n+1)^2}{n(n+2)} = 1 + \frac{1}{n(n+2)}$ together with $\ln(1+x)<x$ for $x>0$, so that $n^2\ln[(n+1)^2/(n(n+2))] < n/(n+2)$ and
\begin{equation}
    \Lambda(n) \;<\; 1 + \frac{n}{n+2} - \frac{4n}{n+1}
    \;=\; -\,\frac{2\,(n^2+2n-1)}{(n+1)(n+2)}.
    \label{eq:Lambda_bound}
\end{equation}
The right-hand side is strictly negative whenever $n^2+2n-1>0$, that is for all $n>\sqrt{2}-1\approx 0.414$, and in particular for every integer $n\geq 1$. Combining~\eqref{eq:rayleigh_exact}--\eqref{eq:Lambda_bound} with the variational principle,
\begin{equation}
    \inf\mathrm{spec}(H) \;\leq\; \frac{n\,\Lambda(n)}{R^2}
    \;<\; -\,\frac{2n\,(n^2+2n-1)}{(n+1)(n+2)}\,\frac{1}{R^2}
    \;<\; 0
    \quad \text{for all } n\geq 1 .
    \label{eq:bound}
\end{equation}
This establishes the existence of at least one negative eigenvalue of $\mathcal{H}^-_0$, hence an exponentially growing off-diagonal perturbation, for the Abelian thick vortex with $n\geq 1$.

Two features of~\eqref{eq:bound} are worth noting. For large $n$ the bound approaches $-2n/R^2 = -B_3(0)$, so the growth rate saturates the peak chromomagnetic field at the core, exactly the scaling of the constant-field Nielsen--Olesen mode. It is also consistent with the elementary limit set by the depth of the well: the kinetic term and the gauge barrier in~\eqref{eq:H_2d} are non-negative, so $\inf\mathrm{spec}(H)\geq\min_r W(r)\geq -2B_3(0)=-4n/R^2$, and indeed $2n(n^2+2n-1)/[(n+1)(n+2)]<4n$ for all $n\geq 1$, the estimate landing at roughly half the available depth at large $n$.

Beyond the Gaussian family, two structural ingredients survive for any profile. The flux normalization $\int_0^\infty B_3\,r\,\mathrm{d}r=n>0$ fixed by~\eqref{eq:total_flux} forces $B_3>0$ on a set of positive measure, so $W(r)=f^2/r^2-2B_3$ is negative somewhere; and regularity forces $f=\mathcal{O}(r^2)$, hence $f^2/r^2\to0$ at the origin, so the barrier cannot cancel the attractive term there. Profiles with $f''(0)>0$ satisfy in addition $W(0)=-2f''(0)<0$.

Neither observation is by itself sufficient. The two-dimensional criterion that any attractive potential of non-positive integral supports a bound state~\cite{Simon:1976un,ReedSimon1978} does not apply, because the $m=0$ barrier behaves as $n^2/r^2$ at large radius and $\int W\,\mathrm{d}^2x$ diverges logarithmically. Nor is $W(0)<0$ automatic: a profile behaving as $f\sim r^4$ near the origin has $f''(0)=0$ and hence $W(0)=0$. What decides the question is the quantitative competition between a barrier of strength $n^2$ and a well of depth $n$, and that is what the exact Rayleigh quotient~\eqref{eq:rayleigh_exact} resolves for the Gaussian family, with explicit constants. The two structural ingredients above make the same conclusion natural for any smooth core, in line with numerical studies of the one-loop vacuum energy in color-magnetic vortex backgrounds, which found such bound states for every profile examined~\cite{Bordag:2002sa}.

A crucial point is that the growing mode cannot change the center charge. Being a bound state of $\mathcal{H}^-_0$, it is localized within the core and leaves the gauge field at large radius untouched, so the holonomy of large loops, and with it the center charge $(-1)^n$, is unchanged. (The integer flux~\eqref{eq:total_flux} is conserved exactly within the Abelian sector, but once off-diagonal components are active only its parity is protected; Section~\ref{sec:endpoint} shows how the flow uses this freedom.) The instability therefore preserves the topological sector and reorganizes the core structure: the vortex evolves from the color-diagonal $\sigma_3$ configuration toward one with active $\sigma_1,\sigma_2$ components. In Cartesian components the growing mode, $\Psi\propto\psi(r)$ with $\Xi=0$, has $a^1_1=-a^2_2\propto\psi(r)$ and all other off-diagonal components zero, up to a global rotation in the $(\sigma_1,\sigma_2)$ plane generated by the residual $\mathrm{U}(1)$. It locks the orientation in color space to the orientation in the transverse plane, which is the precursor to non-Abelian core formation.

A linear instability fixes the direction in which the configuration first moves, not where it ends up. By analogy with the non-Abelian vortices of $\mathrm{SU}(2)$ Yang--Mills--Higgs theory~\cite{Hindmarsh:1985xc} and the BPS non-Abelian vortices of supersymmetric gauge theories with sufficient matter content~\cite{Hanany:2003hp,Auzzi:2003fs,Shifman:2004dr}, one might expect the flow to settle on a core in which all three color components are active and held together by commutator terms. Section~\ref{sec:endpoint} shows that in the transverse plane this expectation fails in an instructive way: pure Yang--Mills theory supplies no scale to hold such a core, and the non-Abelian degrees of freedom serve instead as a channel through which multiply wound vortices unwind.

The analogy with the constant-field Nielsen--Olesen instability~\cite{Nielsen:1978rm,Savvidy:1977as,Ambjorn:1980ms} is now precise. There the off-diagonal vector boson reduces to a charged scalar in a uniform field, organized into Landau levels, and the lowest level gives the exact tachyonic eigenvalue $-gB_3$; here the analogue is the bound state of the localized well $-2B_3(r)$, with binding energy of order the peak field $B_3(0)=2n/R^2$. The difference is topological: a constant chromomagnetic background carries no topological charge and the Nielsen--Olesen instability produces a disordered condensate, whereas the center vortex carries quantized $\mathbb{Z}_2$ holonomy~\eqref{eq:thin_holonomy}, so the instability reorganizes the core while preserving the sector.

\section{The Nonlinear Endpoint}
\label{sec:endpoint}

A word is needed first on the physical meaning of the winding number in $\mathrm{SU}(2)$, since the center is $\mathbb{Z}_2$ and the topological charge of a vortex is $n \bmod 2$. Only the parity of $n$ is protected: configurations with odd $n$ carry the nontrivial center element $-\mathds{1}$ and are genuine center vortices, while even $n$ carries $+\mathds{1}$ and is topologically trivial. In particular $n=1$ is the fundamental center vortex, and $n\geq 2$ describes multiply wound configurations that are not protected against relaxation to lower winding within the same center class. The action~\eqref{eq:action_scaling} makes the stakes explicit: at a given core size, a vortex of winding $n$ costs $n^2$ times as much as a fundamental one.

\subsection{No Static Endpoint}
\label{sec:derrick}

Before asking which configuration the flow approaches, it is worth asking which configurations it could approach. Throughout, the fields depend only on the transverse coordinates, which is a consistent truncation: with $A_3=A_4=0$ and no dependence on $x_3,x_4$ one has $F_{i3}=F_{i4}=0$, so the flow never generates longitudinal components, and their own linearized flow, $\partial_t A_{3,4}=\bar{D}^2A_{3,4}$, is stable. In this setting the action per unit world-sheet area is scale covariant: under $A_\lambda(x)=\lambda A(\lambda x)$ it transforms as $S[A_\lambda]=\lambda^2S[A]$. At a stationary point the derivative with respect to $\lambda$ must vanish, which forces $S=0$, hence $F_{\mu\nu}=0$ and a trivial holonomy. There is therefore no stationary vortex in the transverse plane, Abelian or not, carrying a nontrivial center charge. This is Derrick's argument~\cite{Derrick:1964ww}, and it applies to non-Abelian cores just as much as to the Gaussian one: the flow cannot come to rest on a static non-Abelian vortex, because none exists. The natural candidates for the late stage of the flow are instead the self-similar, expanding solutions $A(x,t)=L(t)^{-1}a\bigl(x/L(t)\bigr)$, whose shape is frozen while the core keeps diluting. The Gaussian vortex of each winding is one of them, and the question becomes which of these the flow selects.

\subsection{Competition with the Dilution}
\label{sec:competition}

The bound~\eqref{eq:bound} is a statement about a frozen core. The background, however, spreads while the mode grows, and since the growth rate is set by the peak field it falls off as $1/R_{\rm eff}^2$, exactly as fast as the core dilutes, so the two effects compete on equal terms at all times. Self-similar variables settle the competition. In
\begin{equation}
    \xi = \frac{x}{L(t)}, \qquad \tau = \ln\frac{L(t)}{R}, \qquad
    L(t)^2 = R^2+4t,
    \label{eq:ss_variables}
\end{equation}
the Gaussian vortex~\eqref{eq:gaussian_solution} is stationary, $f = F(\xi) = n(1-e^{-\xi^2})$. Writing the dangerous helicity component as $\Psi = L^{-1}e^{im\phi}e^{\mu\tau}g(\xi)$, so that it scales like the background potential, the linearized flow~\eqref{eq:Psi_eq} becomes
\begin{equation}
    \mu\,g = \frac12\left[g''+\frac{g'}{\xi}-\frac{(m+F)^2}{\xi^2}\,g\right]
      + \mathcal{B}\,g + \xi g' + g,
    \qquad \mathcal{B} = \frac{F'}{\xi} = 2n\,e^{-\xi^2},
    \label{eq:selfsim}
\end{equation}
the last two terms coming from the dilation. A mode with $\mu>0$ grows relative to the background as $(L/R)^{\mu}$, a power of the core area; one with $\mu<0$ is diluted away faster than the vortex itself. The substitution $g = e^{-\xi^2/2}\chi$ turns~\eqref{eq:selfsim} into $\mu = -E/2$, with $E$ an eigenvalue of the self-adjoint radial operator $K_m = -\partial_\xi^2-\xi^{-1}\partial_\xi+(m+F)^2/\xi^2+\xi^2-2\mathcal{B}$, whose harmonic confinement makes the spectrum discrete.

The ground state of every sector with $m\leq 0$ can be written down in closed form. The function
\begin{equation}
    g_m(\xi) = \xi^{|m|}\exp\!\left[-\xi^2-\frac{n}{2}\,
    \mathrm{Ein}(\xi^2)\right],
    \qquad
    \mathrm{Ein}(z) = \int_0^z\frac{1-e^{-s}}{s}\,\mathrm{d}s,
    \label{eq:exactmode}
\end{equation}
obeys the first-order equation $g_m' = -[(m+F)/\xi+2\xi]\,g_m$. Substituting it into~\eqref{eq:selfsim}, every $\xi$-dependent term cancels by virtue of the identity $F+\mathcal{B}/2 = n$, which is nothing but the statement that the Gaussian is the self-similar Abelian profile, and what remains is
\begin{equation}
    \mu_{n,m} = n-1+m, \qquad m\leq 0 .
    \label{eq:mu_exact}
\end{equation}
Since $g_m$ has no nodes, it is the ground state of its sector, and~\eqref{eq:mu_exact} is the exact growth exponent there. The remaining directions are all less unstable: for $m>0$ the barrier exceeds that of $m=0$, $(m+F)^2-F^2 = m^2+2mF>0$, so $\mu_{n,m}<\mu_{n,0}$; the stable helicity carries $+2\mathcal{B}$ in place of $-2\mathcal{B}$ and lies below it again; and in the Abelian sector each Cartesian component obeys the free heat equation, so $\mu\leq -1$.

For $n\geq 2$, then, the $m=0$ mode grows as $(L/R)^{n-1}$, and the sectors $m=-1,\dots,-(n-2)$ are unstable as well, with smaller exponents. For the fundamental vortex every direction has $\mu\leq 0$, with the $m=0$ mode exactly marginal: the frozen-core instability of Section~\ref{sec:stability} is real, but the dilution cancels it exactly, and the $n=1$ vortex keeps its Abelian form.

\subsection{The Geometry of the Instability}
\label{sec:geometry}

The integers in~\eqref{eq:mu_exact} have a transparent geometric origin. The dangerous mode of sector $m$ is invariant under a rotation combined with a color rotation about $T_3$. Consider the whole class of configurations with this symmetry, and pass to the singular gauge $G_k=\exp(-ik\phi T_3)$ with $k=1-m$. This is an $\mathrm{SO}(3)$ gauge transformation, harmless in pure Yang--Mills theory where every field is adjoint, and it shifts the Abelian profile as $f\to f-k$. In that gauge the fields are independent of $\phi$, $A=A_r(r)\,\mathrm{d}r+\Phi(r)\,\mathrm{d}\phi$, and in radial gauge $A_r=0$ the action per unit world-sheet area becomes
\begin{equation}
    S = \pi\int_0^\infty \frac{|\Phi'(r)|^2}{r}\,\mathrm{d}r,
    \qquad \Phi(r)\in\mathfrak{su}(2)\cong\mathbb{R}^3 .
    \label{eq:curve_action}
\end{equation}
A symmetric configuration is thus a curve in $\mathbb{R}^3$, and its action is a weighted length. Regularity at the origin pins the starting point, $\Phi(0)=-k\hat{e}$, while the holonomy of the circle of radius $r$ is conjugate to $(-1)^k\exp\bigl(2\pi i\,\Phi(r)\bigr)$, so the asymptotic holonomy fixes $|\Phi(\infty)|=|n-k|$, and the flow, which leaves the field at infinity untouched, preserves it. The Abelian vortex is the straight segment $\Phi=(f-k)\hat{e}$, running from $-k\hat{e}$ to $(n-k)\hat{e}$.

When $k<n$ this segment passes through the origin of $\mathbb{R}^3$, where the angular part of $|\Phi'|^2$ carries no weight. It can therefore be swung about the origin and made to end at $-(n-k)\hat{e}$ instead, a radial segment of length $|n-2k|$ rather than $n$. That is an Abelian vortex of winding $2k-n$, in the same center class, at a fraction $(n-2k)^2/n^2$ of the action for a core of the same size. The unstable mode of sector $m$ is the first step of this swing, and it exists precisely when the segment crosses the origin, $k<n$, in agreement with $\mu_{n,m}=n-k$. When $k=n$ the segment ends at the origin and the swing changes nothing, which is the marginal mode. Since the holonomy of the circle of radius $r$ is a function of $\Phi(r)$, the swing has a gauge-invariant meaning as well: the path traced in $\mathrm{SU}(2)$ by the holonomy of a growing circle, which for the Abelian vortex winds $n$ times half-way around a single great circle, is rerouted along a shorter one. The reorganization does not build a new kind of core; it shortens the holonomy path. Repeating the argument, an unstable sector exists as long as $|n|\geq 2$, and the expected endpoint is the Abelian expander of winding $n\bmod 2$: the fundamental vortex for odd $n$, and the vacuum for even $n$.

The full flow is of course free to leave the symmetric sectors, for instance by separating the core into unit vortices, a channel the action also favors, since $n$ separated unit vortices of the same core size cost $\pi n/R^2$ against $\pi n^2/R^2$. Such routes change the path but not the destination. Separated cores keep spreading, and once their width exceeds their separation they are seen from outside as a single core of winding $n$, to which the swing applies; the asymptotic holonomy fixes the center class throughout; and the scaling argument of Section~\ref{sec:derrick} leaves no static configuration along the way. The endpoint, the Abelian expander of winding $n\bmod 2$, is the same whichever channel the flow takes first.

\subsection{Relation to the Flow on Compact Surfaces}
\label{sec:compact}

This picture has a precise antecedent in mathematics. On a compact Riemann surface the Yang--Mills functional is a Morse function in the equivariant sense, its critical points are the reducible connections of constant curvature, labeled by the charge of the reduction, and the index of each grows with that charge~\cite{Atiyah:1982fa}; the gradient flow converges to one of them~\cite{Rade:1992}. The exponents~\eqref{eq:mu_exact} are the planar version of this structure: a vortex of winding $n$ has exactly $n-1$ unstable sectors, $m=0,-1,\dots,-(n-2)$, each a complex direction, in the same way that the index of a reducible critical point counts the charged modes in its lowest Landau level. The plane differs from a compact surface in two respects that matter here. There is no minimum at finite size, so the critical points of Atiyah and Bott are replaced by self-similar expanders, and the boundary at infinity carries the holonomy, which is what fixes the center class and selects winding $n\bmod 2$ as the end of the cascade.

\section{Summary and Conclusions}
\label{sec:conclusions}

We have analyzed the Yang--Mills gradient flow for $\mathrm{SU}(2)$ center vortex configurations, focusing on the interplay between the color-diagonal structure of the thick vortex ansatz and the off-diagonal degrees of freedom it suppresses.

For the thick center vortex $\bar{A}_\mu = f(r,t)\,\partial_\mu\phi\,T_3$, the Yang--Mills flow reduces exactly to the linear diffusion equation $\partial_t f = r\,\partial_r(r^{-1}\partial_r f)$, the relative sign of the first-derivative term reflecting the fact that $f$ is a one-form component rather than a scalar. The bounded fixed points are the constant profiles, and the holonomy condition selects $f_*=n$, the thin vortex. For a Gaussian core the flow is solvable in closed form, $f(r,t) = n[1-e^{-r^2/(R^2+4t)}]$: the width grows as $\sqrt{R^2+4t}$ while the flux $2\pi n$, the regularity condition $f(0,t)=0$ and the asymptotic value $f(\infty,t)=n$ are preserved exactly. The thin vortex is thus a fixed point but not an attractor. Indeed the action of the family, $\pi n^2/R_{\rm eff}^2$, diverges in the thin limit, so that configuration is the maximum of the action rather than its minimum and a steepest-descent flow must move away from it. Linearizing in the color-diagonal sector gives an operator symmetric on the measure $\mathrm{d}r/r$ with quadratic form $\int (g')^2\,\mathrm{d}r/r \geq 0$, hence a purely continuous non-negative spectrum with eigenmodes $rJ_1(\sqrt{\lambda}\,r)$. There is no Abelian tachyonic mode, but no regular Abelian stationary configuration either: the evolution is a scale-dilution runaway.

The picture changes qualitatively when off-diagonal $\sigma_1,\sigma_2$ perturbations are considered. Decomposing fluctuations into the charge-$\pm 1$ basis $a_\mu^\pm = a_\mu^1 \pm ia_\mu^2$ under the background $\mathrm{U}(1)$, and then into helicity eigenstates within the charge-$+$ sector, the linearized flow operator reduces to a pair of self-adjoint radial Schr\"odinger operators $\mathcal{H}_m^\pm$ of the form~\eqref{eq:Hpm}. The $\mathcal{H}_m^+$ sector is stable. The $\mathcal{H}_m^-$ sector contains an attractive chromomagnetic potential $-2B_3(r) = -2f'(r)/r$ that competes with the centrifugal-plus-gauge barrier $(m+f)^2/r^2$. For the $m=0$ mode the barrier vanishes at the origin, since regularity forces $f=\mathcal{O}(r^2)$, so the attractive term dominates near the core whenever $B_3(0)=f''(0)>0$, as it is for the Gaussian.

Whether this well binds is a quantitative question: the gauge barrier carries a factor $n^2$ against the $n$ of the magnetic term and decays only as $n^2/r^2$, so the standard two-dimensional criterion does not apply. We settled it variationally with a Gaussian trial function, retaining every term. For a Gaussian core all three expectation values are available in closed form, including the barrier~\eqref{eq:Vgauge}, and the exact Rayleigh quotient~\eqref{eq:rayleigh_exact} is negative near $\sigma^2\simeq R^2/n$, giving
\begin{equation}
    \inf\mathrm{spec}(\mathcal{H}^-_0)
    < -\,\frac{2n\,(n^2+2n-1)}{(n+1)(n+2)}\,\frac{1}{R^2} \;<\; 0 ,
    \label{eq:central_bound}
\end{equation}
for every $n\geq 1$: frozen at any instant, the Gaussian-core Abelian thick vortex is unstable under off-diagonal perturbations, with a growth rate tending to the peak field $B_3(0)=2n/R^2$ at large $n$. The same mechanism operates for other smooth cores, the flux normalization guaranteeing an attractive region and regularity guaranteeing that the barrier vanishes at the origin, as numerical evidence for a range of profiles confirms~\cite{Bordag:2002sa}. The instability is the inhomogeneous-field analogue of the Nielsen--Olesen instability of the Savvidy vacuum~\cite{Nielsen:1978rm,Savvidy:1977as,Ambjorn:1980ms}, localized within the core on scale $R$.

The frozen instability, however, is only half of the story, because the core spreads while the mode grows and the growth rate falls off exactly as fast as the core dilutes. In self-similar variables the competition is settled exactly. The unstable modes are known in closed form~\eqref{eq:exactmode}, and their growth exponents relative to the background are the integers $\mu_{n,m}=n-1+m$ for $m\leq 0$. Every multiply wound vortex is unstable, with the off-diagonal amplitude growing as $(L/R)^{n-1}$, while for the fundamental vortex the frozen tachyon is cancelled exactly by the dilution and the core keeps its Abelian form.

The nonlinear endpoint follows from two observations. The action in the transverse plane is scale covariant, so no static vortex with a center charge exists, Abelian or not, and the late stage of the flow is described by self-similar expanders. Within each symmetric sector a vortex is a curve in $\mathfrak{su}(2)$ whose action is a weighted length, the Abelian vortex is a straight segment through the origin, and the unstable mode swings it into a shorter one, lowering the winding by two. The holonomy of large loops, and with it the center charge $(-1)^n$, is never touched. The cascade ends at the Abelian expander of winding $n\bmod 2$: the fundamental vortex for odd $n$ and the vacuum for even $n$. Non-Abelian structure is therefore real but transient, the route by which multiply wound cores unwind, and the whole picture is the planar, non-compact counterpart of the Morse theory of Yang--Mills theory on Riemann surfaces~\cite{Atiyah:1982fa,Rade:1992}.

The center vortex mechanism for confinement relies on the topological holonomy of the vortex, which the flow preserves. The quantitative string tension in the semiclassical dilute-vortex-gas picture~\cite{Reinhardt:2001kf,Engelhardt:1999wr} depends in addition on the vortex action, which is determined by the core. Our results say that only fundamental cores survive the flow, and that they survive in Abelian form, so the Abelian core is the right starting point for the vortex action at the classical level, while multiply wound configurations relax to it at a rate fixed by~\eqref{eq:mu_exact}.

On the lattice, whenever the Yang--Mills gradient flow is applied before center projection to smooth UV fluctuations, fundamental vortices keep their Abelian form and their topology throughout. Multiply wound cores do not: their off-diagonal content grows as $(L/R)^{n-1}$, and they unwind once the core has grown by a factor of order $\delta^{-1/(n-1)}$, with $\delta$ the initial off-diagonal admixture. Vortex observables extracted after extensive smoothing therefore measure the fundamental population rather than the full structure present in the original configuration.

The most interesting direction this work opens is finite temperature. Every conclusion above about the endpoint rests on the scale invariance of the transverse plane, and a compact Euclidean time direction breaks it. The time component of the gauge field then behaves as an adjoint scalar whose expectation value is set by the Polyakov loop, which is precisely the ingredient that holds non-Abelian vortex cores together in Yang--Mills--Higgs theories~\cite{Hanany:2003hp,Auzzi:2003fs,Shifman:2004dr}. Above the deconfinement transition the flow may therefore come to rest on a genuinely non-Abelian vortex, the instability of the Abelian core and the scale set by the temperature competing to fix its size, with direct consequences for the percolation picture of deconfinement~\cite{Engelhardt:1999wr,Langfeld:1998cz}. We will return to this question in a separate publication. Two further extensions are natural. In $\mathrm{SU}(N)$ the center is $\mathbb{Z}_N$, the holonomy path lives in a larger group, and the same geometry predicts a richer cascade toward the shortest path in each $N$-ality class. In four dimensions vortex world-sheets are closed and curved, and their intersection points carry fractional topological charge~\cite{Engelhardt:1999xw,Reinhardt:2001kf,Hollwieser:2011uj}; the curvature radius supplies another scale, and the fate of the intersection points under the flow bears directly on the vortex picture of topological charge. Finally, the functional determinant of $\mathcal{H}_m^\pm$ controls the one-loop weight of the vortex. Around the Abelian core the negative mode makes that weight ill-defined, as one-loop studies of color-magnetic vortex backgrounds have also found~\cite{Bordag:2002sa}, and at zero temperature there is no static configuration to expand around; at finite temperature, where a static core can exist, extending existing one-loop analyses of center-vortex configurations~\cite{Diakonov:2002bx} would give a direct semiclassical prediction for the string tension.

\section*{Acknowledgements}
The author would like to thank D. R. Junior for the initial discussions that shaped this work.

\bibliographystyle{unsrt}
\bibliography{vortex_flow}

@article{tHooft:1977nqb,
  author  = {'t Hooft, G.},
  title   = {On the phase transition towards permanent quark confinement},
  journal = {Nucl.\ Phys.\ B},
  volume  = {138},
  pages   = {1--25},
  year    = {1978},
  doi     = {10.1016/0550-3213(78)90153-0}
}

@article{tHooft:1981ht,
  author  = {'t Hooft, G.},
  title   = {Topology of the gauge condition and new confinement phases in
             non-{Abelian} gauge theories},
  journal = {Nucl.\ Phys.\ B},
  volume  = {190},
  pages   = {455--478},
  year    = {1981},
  doi     = {10.1016/0550-3213(81)90442-9}
}

@article{Cornwall:1979hz,
  author  = {Cornwall, J. M.},
  title   = {Quark confinement and vortices in massive gauge-invariant
             {QCD}},
  journal = {Nucl.\ Phys.\ B},
  volume  = {157},
  pages   = {392--412},
  year    = {1979},
  doi     = {10.1016/0550-3213(79)90111-1}
}

@article{DelDebbio:1996mh,
  author  = {Del Debbio, L. and Faber, M. and Greensite, J. and Olejn{\'\i}k, {\v S}.},
  title   = {Center dominance and {$Z(2)$} vortices in {SU(2)} lattice gauge
             theory},
  journal = {Phys.\ Rev.\ D},
  volume  = {55},
  pages   = {2298--2306},
  year    = {1997},
  eprint  = {hep-lat/9610005},
  doi     = {10.1103/PhysRevD.55.2298}
}

@article{Langfeld:1997jx,
  author  = {Langfeld, K. and Reinhardt, H. and Tennert, O.},
  title   = {Confinement and scaling of the vortex vacuum of {SU(2)} lattice
             gauge theory},
  journal = {Phys.\ Lett.\ B},
  volume  = {419},
  pages   = {317--321},
  year    = {1998},
  eprint  = {hep-lat/9710068},
  doi     = {10.1016/S0370-2693(97)01435-4}
}

@article{Faber:2001zs,
  author  = {Faber, M. and Greensite, J. and Olejn{\'\i}k, {\v S}.},
  title   = {Direct {Laplacian} center gauge},
  journal = {JHEP},
  volume  = {11},
  pages   = {053},
  year    = {2001},
  eprint  = {hep-lat/0106017},
  doi     = {10.1088/1126-6708/2001/11/053}
}

@article{deForcrand:1999ms,
  author  = {de Forcrand, Ph. and D'Elia, M.},
  title   = {On the relevance of center vortices to {QCD}},
  journal = {Phys.\ Rev.\ Lett.},
  volume  = {82},
  pages   = {4582--4585},
  year    = {1999},
  eprint  = {hep-lat/9901020},
  doi     = {10.1103/PhysRevLett.82.4582}
}

@article{Engelhardt:1999fd,
  author  = {Engelhardt, M. and Reinhardt, H.},
  title   = {Center vortex model for the infrared sector of
             {Yang--Mills} theory: Confinement and deconfinement},
  journal = {Nucl.\ Phys.\ B},
  volume  = {585},
  pages   = {591--613},
  year    = {2000},
  eprint  = {hep-lat/9912003},
  doi     = {10.1016/S0550-3213(00)00445-4}
}

@article{Engelhardt:1999wr,
  author  = {Engelhardt, M. and Langfeld, K. and Reinhardt, H.
             and Tennert, O.},
  title   = {Deconfinement in {SU(2)} {Yang--Mills} theory as a center
             vortex percolation transition},
  journal = {Phys.\ Rev.\ D},
  volume  = {61},
  pages   = {054504},
  year    = {2000},
  eprint  = {hep-lat/9904004},
  doi     = {10.1103/PhysRevD.61.054504}
}

@article{Langfeld:1998cz,
  author  = {Langfeld, K. and Tennert, O. and Engelhardt, M. and
             Reinhardt, H.},
  title   = {Center vortices of {Yang--Mills} theory at finite temperatures},
  journal = {Phys.\ Lett.\ B},
  volume  = {452},
  pages   = {301--309},
  year    = {1999},
  eprint  = {hep-lat/9805002},
  doi     = {10.1016/S0370-2693(99)00252-X}
}

@article{Greensite:2003bk,
  author  = {Greensite, J.},
  title   = {The confinement problem in lattice gauge theory},
  journal = {Prog.\ Part.\ Nucl.\ Phys.},
  volume  = {51},
  pages   = {1--83},
  year    = {2003},
  eprint  = {hep-lat/0301023},
  doi     = {10.1016/S0146-6410(03)90012-3}
}

@book{Greensite:2016pfc,
  author    = {Greensite, J.},
  title     = {An Introduction to the Confinement Problem},
  series    = {Lecture Notes in Physics},
  volume    = {821},
  publisher = {Springer},
  address   = {Berlin},
  year      = {2011},
  doi       = {10.1007/978-3-642-14382-3}
}

@article{Engelhardt:1999xw,
  author  = {Engelhardt, M. and Reinhardt, H.},
  title   = {Center projection vortices in continuum {Yang--Mills} theory},
  journal = {Nucl.\ Phys.\ B},
  volume  = {567},
  pages   = {249--292},
  year    = {2000},
  eprint  = {hep-th/9907139},
  doi     = {10.1016/S0550-3213(99)00727-0}
}

@article{Reinhardt:2001kf,
  author  = {Reinhardt, H.},
  title   = {Topology of center vortices},
  journal = {Nucl.\ Phys.\ B},
  volume  = {628},
  pages   = {133--166},
  year    = {2002},
  eprint  = {hep-th/0112215},
  doi     = {10.1016/S0550-3213(02)00130-X}
}

@article{deForcrand:2000pg,
  author  = {de Forcrand, Ph. and Pepe, M.},
  title   = {Center vortices and monopoles without lattice {Gribov} copies},
  journal = {Nucl.\ Phys.\ B},
  volume  = {598},
  pages   = {557--577},
  year    = {2001},
  eprint  = {hep-lat/0008016},
  doi     = {10.1016/S0550-3213(01)00009-8}
}

@article{Hollwieser:2011uj,
  author  = {Hollwieser, R. and Faber, M. and Heller, U. M.},
  title   = {Intersections of thick center vortices, {Dirac} eigenmodes and
             fractional topological charge in {SU(2)} lattice gauge theory},
  journal = {JHEP},
  volume  = {06},
  pages   = {052},
  year    = {2011},
  eprint  = {1103.2669},
  doi     = {10.1007/JHEP06(2011)052}
}

@article{Trewartha:2015ida,
  author  = {Trewartha, A. and Kamleh, W. and Leinweber, D.},
  title   = {Evidence that centre vortices underpin dynamical chiral
             symmetry breaking in {SU(3)} gauge theory},
  journal = {Phys.\ Lett.\ B},
  volume  = {747},
  pages   = {373--377},
  year    = {2015},
  eprint  = {1502.06753},
  doi     = {10.1016/j.physletb.2015.06.025}
}

@article{Biddle:2019gke,
  author  = {Biddle, J. C. and Kamleh, W. and Leinweber, D. B.},
  title   = {Visualization of center vortex structure},
  journal = {Phys.\ Rev.\ D},
  volume  = {102},
  pages   = {034504},
  year    = {2020},
  eprint  = {1912.09531},
  doi     = {10.1103/PhysRevD.102.034504}
}

@article{Belavin:1975fg,
  author  = {Belavin, A. A. and Polyakov, A. M. and Schwartz, A. S.
             and Tyupkin, Yu. S.},
  title   = {Pseudoparticle solutions of the {Yang--Mills} equations},
  journal = {Phys.\ Lett.\ B},
  volume  = {59},
  pages   = {85--87},
  year    = {1975},
  doi     = {10.1016/0370-2693(75)90163-X}
}

@article{Callan:1977gz,
  author  = {Callan, C. G. and Dashen, R. F. and Gross, D. J.},
  title   = {Toward a theory of the strong interactions},
  journal = {Phys.\ Rev.\ D},
  volume  = {17},
  pages   = {2717--2763},
  year    = {1978},
  doi     = {10.1103/PhysRevD.17.2717}
}

@article{Narayanan:2006rf,
  author  = {Narayanan, R. and Neuberger, H.},
  title   = {Infinite {$N$} phase transitions in continuum {Wilson} loop
             operators},
  journal = {JHEP},
  volume  = {03},
  pages   = {064},
  year    = {2006},
  eprint  = {hep-th/0601210},
  doi     = {10.1088/1126-6708/2006/03/064}
}

@article{Luscher:2009eq,
  author  = {L{\"u}scher, M.},
  title   = {Trivializing maps, the {Wilson} flow and the {HMC} algorithm},
  journal = {Commun.\ Math.\ Phys.},
  volume  = {293},
  pages   = {899--919},
  year    = {2010},
  eprint  = {0907.5491},
  doi     = {10.1007/s00220-009-0953-7}
}

@article{Luscher:2010iy,
  author  = {L{\"u}scher, M.},
  title   = {Properties and uses of the {Wilson} flow in lattice {QCD}},
  journal = {JHEP},
  volume  = {08},
  pages   = {071},
  year    = {2010},
  eprint  = {1006.4518},
  doi     = {10.1007/JHEP08(2010)071}
}

@article{Luscher:2011bx,
  author  = {L{\"u}scher, M. and Weisz, P.},
  title   = {Perturbative analysis of the gradient flow in non-{Abelian}
             gauge theories},
  journal = {JHEP},
  volume  = {02},
  pages   = {051},
  year    = {2011},
  eprint  = {1101.0963},
  doi     = {10.1007/JHEP02(2011)051}
}

@article{Luscher:2013cpa,
  author  = {L{\"u}scher, M.},
  title   = {Chiral symmetry and the {Yang--Mills} gradient flow},
  journal = {JHEP},
  volume  = {04},
  pages   = {123},
  year    = {2013},
  eprint  = {1302.5246},
  doi     = {10.1007/JHEP04(2013)123}
}

@article{Suzuki:2013gza,
  author  = {Suzuki, H.},
  title   = {Energy--momentum tensor from the {Yang--Mills} gradient flow},
  journal = {Prog.\ Theor.\ Exp.\ Phys.},
  volume  = {2013},
  pages   = {083B03},
  year    = {2013},
  eprint  = {1304.0533},
  doi     = {10.1093/ptep/ptt059}
}

@article{Luscher:2013vga,
  author  = {L{\"u}scher, M.},
  title   = {Future applications of the {Yang--Mills} gradient flow in
             lattice {QCD}},
  journal = {PoS},
  volume  = {LATTICE2013},
  pages   = {016},
  year    = {2014},
  eprint  = {1308.5598},
  doi     = {10.22323/1.187.0016}
}

@article{Borsanyi:2012zs,
  author  = {Borsanyi, S. and D{\"u}rr, S. and Fodor, Z. and Hoelbling, C.
             and Katz, S. D. and Krieg, S. and Kurth, T. and Lellouch, L.
             and Lippert, T. and McNeile, C. and Szab{\'o}, K. K.},
  title   = {High-precision scale setting in lattice {QCD}},
  journal = {JHEP},
  volume  = {09},
  pages   = {010},
  year    = {2012},
  eprint  = {1203.4469},
  doi     = {10.1007/JHEP09(2012)010}
}

@article{Bonati:2014tqa,
  author  = {Bonati, C. and D'Elia, M.},
  title   = {Comparison of the gradient flow with cooling in {SU(3)}
             pure gauge theory},
  journal = {Phys.\ Rev.\ D},
  volume  = {89},
  pages   = {105005},
  year    = {2014},
  eprint  = {1401.2441},
  doi     = {10.1103/PhysRevD.89.105005}
}

@article{Alexandrou:2015yba,
  author  = {Alexandrou, C. and Athenodorou, A. and Jansen, K.},
  title   = {Topological charge using cooling and the gradient flow},
  journal = {Phys.\ Rev.\ D},
  volume  = {92},
  pages   = {125014},
  year    = {2015},
  eprint  = {1509.04259},
  doi     = {10.1103/PhysRevD.92.125014}
}

@article{Nielsen:1978rm,
  author  = {Nielsen, N. K. and Olesen, P.},
  title   = {An unstable {Yang--Mills} field mode},
  journal = {Nucl.\ Phys.\ B},
  volume  = {144},
  pages   = {376--396},
  year    = {1978},
  doi     = {10.1016/0550-3213(78)90377-2}
}

@article{Savvidy:1977as,
  author  = {Savvidy, G. K.},
  title   = {Infrared instability of the vacuum state of gauge theories
             and asymptotic freedom},
  journal = {Phys.\ Lett.\ B},
  volume  = {71},
  pages   = {133--134},
  year    = {1977},
  doi     = {10.1016/0370-2693(77)90759-6}
}

@article{Ambjorn:1980ms,
  author  = {Ambj{\o}rn, J. and Olesen, P.},
  title   = {On the formation of a random color magnetic quantum liquid
             in {QCD}},
  journal = {Nucl.\ Phys.\ B},
  volume  = {170},
  pages   = {60--78},
  year    = {1980},
  doi     = {10.1016/0550-3213(80)90476-9}
}

@article{Bordag:2002sa,
  author  = {Bordag, M.},
  title   = {On the vacuum energy of a color magnetic vortex},
  journal = {Phys.\ Rev.\ D},
  volume  = {67},
  pages   = {065001},
  year    = {2003},
  eprint  = {hep-th/0211080},
  doi     = {10.1103/PhysRevD.67.065001}
}

@article{Diakonov:2002bx,
  author  = {Diakonov, D. and Maul, M.},
  title   = {Center vortex solutions of the {Yang--Mills} effective action
             in three and four dimensions},
  journal = {Phys.\ Rev.\ D},
  volume  = {66},
  pages   = {096004},
  year    = {2002},
  eprint  = {hep-lat/0204012},
  doi     = {10.1103/PhysRevD.66.096004}
}

@article{Hindmarsh:1985xc,
  author  = {Hindmarsh, M. and Kibble, T. W. B.},
  title   = {Beads on strings},
  journal = {Phys.\ Rev.\ Lett.},
  volume  = {55},
  pages   = {2398--2400},
  year    = {1985},
  doi     = {10.1103/PhysRevLett.55.2398}
}

@article{Hanany:2003hp,
  author  = {Hanany, A. and Tong, D.},
  title   = {Vortex strings and four-dimensional gauge dynamics},
  journal = {JHEP},
  volume  = {04},
  pages   = {066},
  year    = {2004},
  eprint  = {hep-th/0403158},
  doi     = {10.1088/1126-6708/2004/04/066}
}

@article{Auzzi:2003fs,
  author  = {Auzzi, R. and Bolognesi, S. and Evslin, J. and Konishi, K.
             and Yung, A.},
  title   = {Non-{Abelian} superconductors: Vortices and confinement
             in {$\mathcal{N}=2$} {SQCD}},
  journal = {Nucl.\ Phys.\ B},
  volume  = {673},
  pages   = {187--216},
  year    = {2003},
  eprint  = {hep-th/0307287},
  doi     = {10.1016/j.nuclphysb.2003.09.029}
}

@article{Shifman:2004dr,
  author  = {Shifman, M. and Yung, A.},
  title   = {Non-{Abelian} string junctions as confined monopoles},
  journal = {Phys.\ Rev.\ D},
  volume  = {70},
  pages   = {045004},
  year    = {2004},
  eprint  = {hep-th/0403149},
  doi     = {10.1103/PhysRevD.70.045004}
}

@article{Atiyah:1982fa,
  author  = {Atiyah, M. F. and Bott, R.},
  title   = {The {Yang--Mills} equations over {Riemann} surfaces},
  journal = {Phil.\ Trans.\ Roy.\ Soc.\ Lond.\ A},
  volume  = {308},
  pages   = {523--615},
  year    = {1983},
  doi     = {10.1098/rsta.1983.0017}
}

@article{Rade:1992,
  author  = {R{\aa}de, J.},
  title   = {On the {Yang--Mills} heat equation in two and three dimensions},
  journal = {J.\ Reine Angew.\ Math.},
  volume  = {431},
  pages   = {123--164},
  year    = {1992},
  doi     = {10.1515/crll.1992.431.123}
}

@article{Derrick:1964ww,
  author  = {Derrick, G. H.},
  title   = {Comments on nonlinear wave equations as models for elementary
             particles},
  journal = {J.\ Math.\ Phys.},
  volume  = {5},
  pages   = {1252--1254},
  year    = {1964},
  doi     = {10.1063/1.1704233}
}

@article{Simon:1976un,
  author  = {Simon, B.},
  title   = {The bound state of weakly coupled {Schr\"odinger} operators
             in one and two dimensions},
  journal = {Ann.\ Phys.},
  volume  = {97},
  pages   = {279--288},
  year    = {1976},
  doi     = {10.1016/0003-4916(76)90038-5}
}

@book{ReedSimon1978,
  author    = {Reed, M. and Simon, B.},
  title     = {Methods of Modern Mathematical Physics {IV}: Analysis of
               Operators},
  publisher = {Academic Press},
  address   = {New York},
  year      = {1978}
}

@book{Berline:2003,
  author    = {Berline, N. and Getzler, E. and Vergne, M.},
  title     = {Heat Kernels and {Dirac} Operators},
  publisher = {Springer},
  address   = {Berlin},
  year      = {2004}
}

@book{Grigoryan:2009,
  author    = {Grigor'yan, A.},
  title     = {Heat Kernel and Analysis on Manifolds},
  publisher = {American Mathematical Society},
  address   = {Providence, RI},
  year      = {2009}
}

\end{document}